\documentclass[twocolumn,showpacs,aps]{revtex4}

\usepackage{graphicx}
\usepackage{dcolumn}
\usepackage{bm}

\renewcommand{\vec}[1]{\mbox{\boldmath $#1$}}

\begin{document}

\preprint{}

\title{
No-recoil approximation to knock-on exchange potential \\ 
in the double folding model for heavy-ion collisions} 

\author{K. Hagino}
\affiliation{
Department of Physics, Tohoku University, Sendai 980-8578, Japan}

\author{T. Takehi}
\affiliation{
Department of Physics, Tohoku University, Sendai 980-8578, Japan}

\author{N. Takigawa}
\affiliation{
Department of Physics, Tohoku University, Sendai 980-8578, Japan}

\date{\today}

\begin{abstract}
We propose the no-recoil approximation, which is valid for 
heavy systems, for a double folding nucleus-nucleus potential. 
With this approximation, the non-local 
knock-on exchange contribution becomes a local form. 
We discuss the applicability of this approximation 
for the elastic scattering of 
$^6$Li + $^{40}$Ca system. 
We find that, for this system and heavier, 
the no-recoil approximation works as good as another 
widely used local approximation which employs 
a local plane wave for the relative 
motion between the colliding nuclei. 
We also compare the results of the no-recoil calculations with those
of the zero-range approximation often used to handle the knock-on exchange
effect.  
\end{abstract}


\maketitle

The double folding model has been widely used 
to describe the real part of optical potential for heavy-ion 
collisions \cite{SL79,BS97,KS00}. 
The direct part of the double folding potential is constructed 
by convoluting an effective nucleon-nucleon interaction 
with the ground state density distributions of the projectile and 
target nuclei. 
In the double folding model, 
the exchange contribution originating from the antisymmetrization 
of the total wave function of the system is customarily
taken into account 
simply through the single nucleon knock-on exchange term. 
The exchange term leads to a non-local potential. 
Since it is cumbersome to handle the resultant integro-differential equation, 
a local approximation has usually been employed. 
In the past, many calculations have been performed along this line 
by introducing a pseudo zero-range nucleon-nucleon 
interaction to mock up the knock-on exchange 
effect \cite{SL79,BS97,LO75}. 
The strength of the pseudo interaction has been 
tuned so as to reproduce exact results of the 
integro-differential equation 
for proton scattering 
from various target nuclei at several incident energies\cite{LO75}. 
This approach, in conjunction with 
the (density dependent) Michigan-three-range Yukawa (M3Y) 
interaction\cite{BBML77,KBLS84}, 
has successfully accounted for observed elastic and inelastic 
scattering for many colliding systems \cite{SL79,BS97}. 

Recently, a more consistent treatment for the exchange term 
has also been considered\cite{KS00,K88,KOB94,KSO97,K01}. 
This approach obtains 
a local potential by employing a local approximation 
to the momentum operator (local momentum approximation) \cite{S75,SM79}. 
Since the local momentum depends explicitly 
on the potential itself, there arises 
the self-consistency problem, which however can be solved 
iteratively. 
Since the exchange potential is directly constructed 
from a given nucleon-nucleon interaction of finite range, 
this approach is more 
favorable than the zero range approximation. In fact, the finite range 
treatment for the exchange term has enjoyed a success in reproducing 
the experimental angular distributions for light heavy-ion scattering 
where the zero range approximation fails\cite{K88,KOB94,KSOK02}. 

Despite its success, however, there is a potential difficulty in this 
approach. That is, the iterative procedure for the self-consistent 
problem may not work 
in the classically forbidden region, where the local momentum 
is imaginary. Although the frozen density approximation 
used in the double folding model 
could be questionable at these low energies, 
one may still attempt to construct 
a nucleus-nucleus potential with the double folding procedure. 

The aim of this paper is to propose an alternative local approximation
to the 
knock-on exchange term in the double folding model, which is applicable even 
in the classically forbidden region. 
To this end, we exploit the fact that the non-locality 
of the exchange potential arises from the recoil effect and thus 
its range is not large for heavy systems \cite{BS97,GPR76}. 
We simply ignore the recoil effect (i.e., introduce the 
no-recoil approximation), and obtain 
a local nucleus-nucleus potential. As in the local momentum 
approximation, the only ingredients needed in our approach are a 
nucleon-nucleon 
interaction and the one-body density matrices of the colliding nuclei. 
Our approach is thus complementary to the local momentum 
approximation, which is valid even for light systems but may not work 
at very low energies, especially at energies below the Coulomb
barrier. 
A similar no-recoil approximation has been
discussed in Refs. \cite{GPR76,KLT91}, as well as for heavy-ion transfer 
reactions in Ref. \cite{BW91}. Here, we systematically investigate the 
applicability of the no-recoil approximation by comparing to the exact 
result as well as 
to the result in the local momentum approximation. 

We begin with the Schr\"odinger equation 
based on the double folding model 
for the relative motion 
between the colliding nuclei,  
\begin{equation}
\left[-\frac{\hbar^2}{2\mu}\nabla^2+V_d(r)+V_C(r)-iW(r)-E\right]
\psi(\vec{r})
+[V_{\rm ex}\psi](\vec{r})=0, 
\end{equation}
where $\mu$ and $V_C$ are the reduced mass and 
the Coulomb potential, respectively, 
and $-iW$ is 
the imaginary potential which simulates the inelastic and 
fusion processes. $V_d$ is the direct contribution of the 
double folding potential given by \cite{SL79,BS97}
\begin{equation}
V_d(r)= 
\int d\vec{r}_Pd\vec{r}_T\,
\rho_P(\vec{r}_P)\rho_T(\vec{r}_T)\,
v(\vec{r}_T-\vec{r}_P-\vec{r}),
\label{direct}
\end{equation}
while the exchange part is given by \cite{L78}
\begin{eqnarray}
[V_{\rm ex}\psi](\vec{r})&=&
\int d\vec{r}_Pd\vec{r}_T\,
\rho_T(\vec{r}_T-\vec{s},\vec{r}_T)\,
\rho_P(\vec{r}_P+\vec{s},\vec{r}_P) \nonumber \\
&&\times v(\vec{s})\,\psi(\vec{r}+\alpha\vec{s}), 
\label{exchange}
\end{eqnarray}
where $\vec{s}=\vec{r}_T-\vec{r}_P-\vec{r}$ and $\alpha=
(A_P+A_T)/A_PA_T=1/A_P+1/A_T$. Here, 
$v(\vec{s})$ is an effective nucleon-nucleon
interaction. 
$\rho_T$ and $\rho_P$ are 
the one-body density matrix for the target and projectile nuclei, 
respectively, and $\rho_i(\vec{r})$ in Eq. (\ref{direct}) is their
diagonal component ($i$=P or T). 
In order to evaluate those density matrices, 
we use the local density approximation 
\cite{KOB94,L78,CB78}, 
\begin{equation}
\rho(\bar{\vec{r}}+\vec{s}/2,
\bar{\vec{r}}-\vec{s}/2)
\sim \rho(\bar{\vec{r}})\,\hat{j}_1(k_F(\bar{\vec{r}})s), 
\end{equation}
where $\hat{j}_1(x)=3(\sin x-x\cos x)/x^3$, and 
evaluate the local Fermi momentum $k_F(\vec{r})$ 
in the extended Thomas-Fermi approximation. 

One can obtain a local approximation to Eq. (\ref{exchange}) 
by noticing 
\begin{equation}
\psi(\vec{r}+\alpha\vec{s}) = 
e^{i\alpha \vec{s}\cdot \hat{\vec{p}}/\hbar}\,
\psi(\vec{r}),  
\end{equation}
and evaluating the momentum operator 
$\hat{\vec{p}}$ in the local WKB approximation 
$\vec{k}(r)\hbar$ ({\it local momentum approximation} 
\cite{KS00,K88,KOB94,KSO97,K01,S75,SM79}). 
This yields a local exchange potential 
\begin{equation}
[V_{\rm ex}\psi](\vec{r})=V_{\rm ex}(r)\,\psi(\vec{r})
\end{equation}
with 
\begin{eqnarray}
V_{\rm ex}(r)&=&
\int d\vec{r}_Pd\vec{r}_T\,
\rho_T(\vec{r}_T-\vec{s},\vec{r}_T)\,
\rho_P(\vec{r}_P+\vec{s},\vec{r}_P) 
\nonumber \\
&&\times v(\vec{s})\,e^{i\alpha \vec{k}(r)\cdot\vec{s}}, 
\label{localm}
\end{eqnarray}
where the magnitude of the local momentum is given by 
\begin{equation}
k(r)=\sqrt{\frac{2\mu}{\hbar^2}\,\left[
E-V_d(r)-V_C(r)-V_{\rm ex}(r)\right]}. 
\label{localK}
\end{equation}
Notice that the local momentum $k(r)$ has to be determined 
consistently to the exchange potential $V_{\rm ex}$, as it appears 
both on the right and left hand sides of Eq. (\ref{localm}). 
One can also obtain the same 
expression for the exchange potential (\ref{localm}) 
by constructing the trivially equivalent local potential for 
Eq. (\ref{exchange}) 
and approximating the relative wave functions with those in the WKB 
approximation, that is, $\psi(\vec{r})\sim 
e^{i\vec{k}(r)\cdot\vec{r}}/\sqrt{k(r)}$. 

A further simplification for the exchange term can be 
achieved for heavy systems. 
To this end, 
we remark that 
$\alpha$ in Eq. (\ref{exchange}) arises from the variation of 
center of mass as a consequence of the exchange of nucleons between the 
projectile
and target nuclei. It is nothing more than the recoil effect due to the 
nucleon 
exchange \cite{BS97,GPR76}, and may be neglected for heavy systems. 
For instance, the value of $\alpha$ is 0.0673 and 0.192 for 
$^{16}$O + $^{208}$Pb and $^{6}$Li + $^{40}$Ca, 
respectively. 
If one neglects $\alpha$ in Eqs. (\ref{exchange}) or 
(\ref{localm}), the exchange
potential $V_{\rm ex}$ is simply given by, 
\begin{equation}
V_{\rm ex}(r)=
\int d\vec{r}_Pd\vec{r}_T\,
\rho_T(\vec{r}_T-\vec{s},\vec{r}_T)\,
\rho_P(\vec{r}_P+\vec{s},\vec{r}_P) v(\vec{s}).
\label{exchange2}
\end{equation}
We call this approximation the {\it no-recoil approximation}. 
Notice that the self-consistency problem is not involved in this 
approximation, in contrast to the local momentum approximation. 

Let us now investigate numerically the applicability of the 
no-recoil approximation. 
For this purpose, we choose the $^6$Li + $^{40}$Ca system. 
We use a version of the density-dependent M3Y (DDM3Y) interaction, 
CDM3Y6\cite{KSO97}, 
as the nucleon-nucleon interaction, $v$. It is given by 
\begin{equation}
v(r)=F_d(\rho)\left[11061.625\,\frac{e^{-4r}}{4r}-2537.5\,
\frac{e^{-2.5r}}{2.5r}\right]~
~~~({\rm MeV}), 
\end{equation}
for the direct part, (\ref{direct}), and 
\begin{eqnarray}
v(r)&=&F_{\rm ex}(\rho) 
\left[-1524.25\,\frac{e^{-4r}}{4r}-518.75\,\frac{e^{-2.5r}}{2.5r}\right.
\nonumber \\
&&\left.
-7.8474\,\frac{e^{-0.7072r}}{0.7072r}
\right]~
~~~({\rm MeV}), 
\label{vexchange}
\end{eqnarray}
for the exchange part, (\ref{exchange}). 
Here, the length is in the unit of fm, and the density 
dependent strength is given by 
\begin{equation}
F(\rho)=C[1+\alpha e^{-\beta\rho}-\gamma\rho]
\end{equation}
with $\rho=\rho_P(\vec{r}_P)+\rho_T(\vec{r}_T)$ and 
$\rho=\rho_P(\vec{r}_P+\vec{s}/2)+\rho_T(\vec{r}_T-\vec{s}/2)$ for
the direct and the exchange contributions, $F_d$ and $F_{\rm ex}$,
respectively. 
The value for $C$, $\alpha$, $\beta$, and $\gamma$ can be found in 
Ref. \cite{KSO97}.
We assume that
the imaginary potential $W$
is proportional to the
double folding potential with the knock-on exchange term estimated
in the zero-range approximation.
For the projectile and target densities, we use the same densities as in
Refs. \cite{SL79} and \cite{FS85}, respectively. 
The normalization factor is set to be unity for all the 
calculations reported below. 
The barrier height thus obtained is 8.44 MeV in the no-recoil
approximation, while it is 8.45, 8.44, and 8.50 MeV at 
$E_{\rm lab}$=30, 50.6, and 156 MeV, respectively, in the local
momentum approximation. 


\begin{figure}
\includegraphics[scale=0.4,clip]{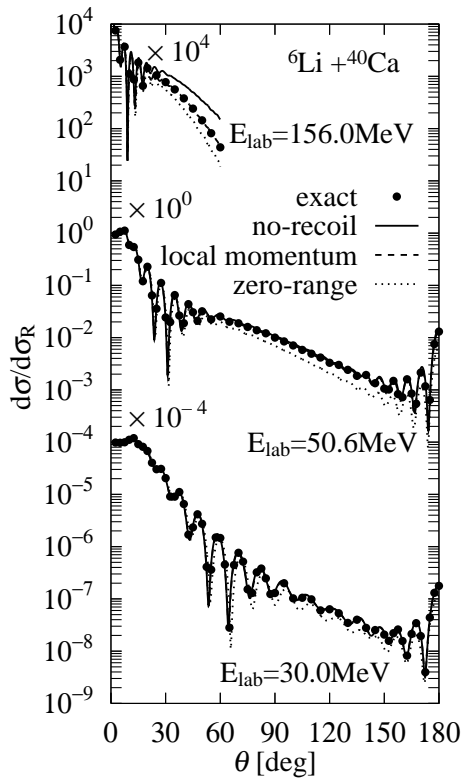}
\caption{
The angular distribution of elastic 
$^6$Li + $^{40}$Ca  
scattering at $E_{\rm lab}$=156.0, 50.6 and 30.0 MeV 
obtained with several methods. 
The filled circles are the exact results of 
the integro-differential equation with the full non-local potential. 
The solid, the dashed, and the dotted lines are obtained in the no-recoil, 
the local momentum, and the zero-range approximations, respectively.
}
\end{figure}

Figure 1 shows the angular distribution 
of elastic $^6$Li + $^{40}$Ca scattering 
at $E_{\rm lab}$=156.0, 50.6 and 30.0 MeV 
as indicated in the figure. 
The filled circles are the exact results of the
integro-differential equation, which fully retains 
the non-locality of the exchange potential. The solid 
and dashed 
lines are obtained in the present no-recoil approximation 
and in the 
local momentum approximations, respectively. 
For comparison, the figure also shows the results of the 
zero-range approximation 
(see the dotted line), that is 
obtained by replacing the nucleon-nucleon potential for
the exchange term given by Eq. (\ref{vexchange}) with 
\begin{equation}
v(\vec{r})=F_{\rm ex}(\rho)\cdot J(E)\delta(\vec{r})
\end{equation}
with $J(E)=-590 (1-0.002 \,E_{\rm lab}/A_P)$ (MeV$\cdot$fm$^{3}$) 
\cite{SL79,BS97,LO75}. 
The strength was tailored particularly for the proton scattering, but has been
used for heavy-ion scattering as well by introducing the dependence on  
the incident energy per projectile nucleon. 

The figure shows that the no-recoil approximation 
leads to similar results as 
the local momentum approximation 
for this system at the lowest two energies. 
They well reproduce the exact results. 
We have confirmed that this is the case 
also for heavier system 
such as 
$^{16}$O + $^{208}$Pb, as it is expected. 
In contrast, 
we observe significant difference between the exact results 
and the results of the zero range approximation. 

At the highest energy, $E_{\rm lab}$=156.0 MeV, 
the no-recoil approximation does not work well. 
This is due to that the local momentum $k(r)$ is relatively large at this
high energy, and the exponent in Eq. (\ref{localm}) 
cannot be neglected even if the value of $\alpha$ itself is small. 
For a lighter system, $\alpha$+$^{90}$Zr, where $\alpha=0.261$ and the
barrier height is around 11.7 MeV, 
we find that the no-recoil approximation does not work even at 
$E_{\rm lab}$=40.0 MeV. 
It is thus clear that, in order for the no-recoil approximation 
to work well, 
the inverse reduced mass $\alpha$ needs to be small and at the same 
time the bombarding energy has to be relatively low. 



In summary, we proposed the no-recoil approximation for the double 
folding model. It neglects the recoil effect due to the knock-on 
exchange of nucleons between the projectile and target nuclei. 
The resultant exchange potential has a simple local form. 
We examined its applicability for heavy-ion reactions by studying the 
angular distribution of elastic 
$^6$Li + $^{40}$Ca scattering. 
We found that the no-recoil 
approximation reproduces reasonably well 
the exact results with the full non-local exchange potential, and 
works as good as the local momentum approximation for 
this system unless the bombarding energy is much above the 
Coulomb barrier. 
The performance of the no-recoil approximation 
improves for heavier systems. 
The zero range approximation, one the other hand, does not 
reproduce well the results of original non-local potential. 
Since the no-recoil approximation 
does not involve the iterative procedure, it is much simpler than the 
local momentum approximation. 
We thus advocate the use of no-recoil approximation 
in analyzing heavy-ion scattering. 

In the double folding model, the exchange effect has been conventionally 
taken into account only through the knock-on exchange term. 
This is reasonable for peripheral collisions, since 
the knock-on exchange has the longest range among other exchange terms 
\cite{AH82}. However, it is not obvious at all whether other exchange 
terms are negligible when the potential in the inner region 
plays a role, such as in rainbow scattering or in fusion reactions. 
In this connection, we mention that 
a similar idea as in the no-recoil approximation proposed in this
paper enables us 
to follow the idea of the 
resonating group method (RGM), that fully incorporates the
exchange effect, relatively easily 
even for heavy systems (see Ref. \cite{F74} for an early attempt). 
We will report on 
such studies in a separate paper\cite{TTH06}. 

\bigskip
We thank T. Wada for useful discussions, and the Australian National
University for their warm hospitality where this work was completed. 
We also thank discussions with the members of the Japan-Australia Cooperative
Scientific Program ``Dynamics of Nuclear Fusion: Evolution Through a
Complex Multi-Dimensional Landscape''.
This work was supported by the Grant-in-Aid for Scientific Research,
Contract No. 16740139 from the Japanese Ministry of Education,
Culture, Sports, Science, and Technology.

\end{document}